# Latent periodicity of serine-threonine and tyrosine protein kinases and another protein families.

Andrew A. Laskin[1,2,] Nikolay A. Kudryashov[2], Konstantin G. Skryabin[1,*], Eugene V. Korotkov[1,2].

[1] *Bioengineering Center of Russian Academy of Sciences, Prospect 60-tya Oktyabrya, 7/1, 117312, Moscow, Russia.*

[2] *Moscow Physical Engineering Institute, Kashirskoe shosse 31, 115409, Moscow, Russia.*

\* to whom correspondence should be addressed
email:bioinf@yandex.ru

Abstract

We identified latent periodicity in catalytic domains of approximately 85% of serine/threonine and tyrosine protein kinases. Similar results were obtained for other 22 protein domains. We also designed the method of noise decomposition, which is aimed to distinguish between different periodicity types of the same period length. The method is to be used in conjunction with the cyclic profile alignment, and this combination is able to reveal structure-related or function-related patterns of latent periodicity. Possible origins of the periodic structure of protein kinase active sites are discussed. Summarizing, we presume that latent periodicity is the common property of many catalytic protein domains.

## 1. Introduction

The development of mathematical techniques for investigation of symbolic sequences is now acquiring ever-growing importance, since large amounts of genetic information are being gathered [1-4]. Which information can be extracted from symbolic sequences using today's mathematical achievements? Answer to this question determines the ability to extract biologically significant knowledge from genetic texts, the understanding of gene evolution processes and evolutionary rearrangements of genomes, and also the ability to create a dynamic model of cell's genetic regulation and artificial proteins with predefined features.

One of the ways to investigate features of a symbolic sequence is studying its periodicity. The investigation of periodicity has reasonable biological meaning because multiple duplications of DNA sequence fragments followed by subsequent substitutions, insertions and deletions of symbols could serve as the ground for evolution of genes and genomes. The discovery of periodicity in active centers of enzymes could witness that, in the past, genes could be built up by simple repeating of certain relatively short DNA fragments. We may also suppose that such structure of protein active sites could mean possible participation of latent amino acid periodicity in choice and stabilization of the proper conformation of protein globule.

The techniques of dynamic programming [5-11] or Fourier transformaion [12-21] are commonly utilized for identification of periodicity. We had developed our own mathematical approach to searching for periodicity, which is based on Information Decomposition (ID) of symbolic sequences [22-26]. The main idea of this approach is that information content of any symbolic sequence could be decomposed into mutually nonoverlapping constituents. Each of the constituents represents the mutual information between the investigated symbolic sequence and the artificial periodic sequence with some period length. The interdependence of mutual information and period length may be presented in the form of spectral graph that resembles Fourier power spectra, but it has substantially different properties [26]. This decomposition allows us to eliminate the shortcomings peculiar to dynamic programming and Fourier transformation, and it allows us to detect so-called latent periodicity, that is, the periodicity that other techniques are powerless to detect.

However, like Fourier-based techniques, the method of information decomposition (in its current form) is not capable of finding statistically significant latent periodicity in presence of many insertions and deletions. This may lead us to the conclusion that substantial part of latent periodicity occurrences in genetic texts remains unseen by information decomposition-based techniques as well as by other known methods. The simplest method of searching for latent periodicity with insertions and deletions is the combination of information decomposition and modified profile analysis. In this combination, information decomposition can serve as the method that detects latent periodicity in some amino acid sequences and creates the periodicity matrix [26], which can be used to determine the weights for each amino acid at each position in the period. Then modified profile analysis allows us to identify periodicity of corresponding type (defined by cyclic position-weight matrix we have just constructed) in all the primary sequence data bank (such as Swiss-Prot) with possible insertions and deletions, and search results can be used to modify the profile matrix for increased sensitivity and specificity of the search.

The first goal of this publication is to present the method of noise decomposition. For many sequence families, perfect tandem periodicity is disrupted with indels; the cyclic alignment thereby is to expand our possibilities of identifying latent periodicities to the cases where sufficiently short indels are present. The method also allows us to distinguish between different periodicity types of the same period length because sometimes there are nearby but distinct types of periodicity. In this paper, we demonstrate that our decomposition technique is able to distinguish two latent repeat profiles as close as those present in serine/threonine and tyrosine kinases.

The second goal is to reveal the prevalence of latent periodicity in protein kinases. Latent periodicity we previously identified in catalytic domains of 7 protein kinases turned out to be more common property of these proteins than one would expect. To achieve this, we applied modified iterative circular profile analysis and the method of noise decomposition; our efforts resulted in certain modification of the initial position-weight matrix and identification of latent periodicity in active sites of 1215 protein kinases presented in Swiss-Prot data bank. The data we gathered witness that latent periodicity is a property of at least a great majority of eukaryotic protein kinases.

The third goal of this publication is to show that there is latent periodicity in a number of different protein families. For this purpose, we applied the ID and noise decomposition methods to investigate some selected protein families from Swiss-prot data bank. We found out that amino acid sequences of many protein domains have latent periodicity with various period lengths. We discuss these results and propose that latent periodicity could reflect the origin of proteins from manifold ancient tandem duplications.

## 2. Methods and algorithms.

Let us define at first, which kinds of periodicity we may call latent. Generally, the latent periodicity is periodicity that is hard to identify with proper level of statistical significance using internal homology search techniques. The homology between periods (repeats) is often determined using amino acid similarity matrices, such as PAM or BLOSUM [8-11], where the weights for similar amino acids are higher than those for dissimilar ones, or using the autocorrelation function algorithm [14].

Let us consider a set of sequences $S^1, S^2, ..., S^N$ of equal length $L$. We want to evaluate the overall similarity between these sequences; to do it, let us construct their (indel-free) multiple alignment:

$$\begin{array}{cccc} S^1_1 & S^1_2 & ... & S^1_L \\ S^2_1 & S^2_2 & ... & S^2_L \\ ... & ... & ... & ... \\ S^N_1 & S^N_2 & ... & S^N_L \end{array}$$

The total weight of this multiple alignment is generally a sum of position weights:

$$W = \sum_{i=1}^{L} W_i . \qquad (1)$$

Here $W_i \equiv W_i(S^1_i, ..., S^N_i)$ is some function that designates the degree of similarity between amino acids $S^1_i, ..., S^N_i$. Traditional homology search techniques calculate this quantity via pair-wise amino acid affinities:

$$W_i = \sum_\alpha \sum_{\beta > \alpha} M(S^\alpha_i, S^\beta_i) , \qquad (2)$$

where $M$ is some amino acid affinity matrix, such as PAM or BLOSUM. This expression may be rewritten in the form of sum by amino acid types:

$$W_i = \frac{1}{2} \sum_{j,k} n_{i,j}(n_{i,k} - \delta^k_j) M(j,k) , \qquad (3)$$

where $j$ and $k$ are amino acid types and $n_{i,j}$ are amino acid frequencies, i.e. the numbers of amino acids of type $j$ at position $i$. In [22-26], we proposed another measure of similarity, which is based on concepts of information theory and called "information content":

$$W_i = \sum_j \frac{n_{i,j}}{Nf_j} \ln \frac{n_{i,j}}{Nf_j} , \qquad (4)$$

where $f_j$ is the frequency of occurrence of amino acids of type $j$ in the whole set of sequences. These measures are undoubtedly different, thus an alignment may score high using information-theoretic measure while scoring low using homology-based measure, and vice versa. But the phrase "high-scoring" does not mean anything, especially when comparing weights calculated with different measures. We have to make sure that the obtained value of $W$ is far above those calculated with sets of random, unrelated sequences. To do it, we shuffle initial sequences and calculate either p-value or Z-value of obtained alignment; low p-values, or high Z-values, witness for significant similarity between sequences $S^1, S^2, ..., S^N$. One usually sets up some threshold value, which exceeding is believed to mean that the similarity is not arisen by chance.

When $S^1, S^2, ..., S^N$ are consecutive equal-length slices of the sequence under investigation, we may say that significant similarity between them means significant periodicity in this sequence. As we said before, different similarity measures result in different weights and different significance values; in some cases, periodicity of a sequence may be apparent from information-theoretical viewpoint while omitted by homology searches. In our studies, we call this effect latent periodicity. In our previous studies [22-26], we have shown that such latent periodicity is present in many sequences of biological importance.

Table 1. The set of amino acids used for generation of latently-periodic sequence with 7 aa long period. At each position, the frequencies of mentioned amino acids were chosen equal.

| Position of the period | 1 | 2 | 3 | 4 | 5 | 6 | 7 |
|---|---|---|---|---|---|---|---|
| Set of the amino acids in the position | ARND CQEZ XWYB | WYVB ZXGK NAIL | GHIL KMFE CQAR N | KMFP STWC QHI | ARND CQIL GZS | GHIL NDCD ZS | EGHI LCQE GS |

Let us illustrate this with an example. Assume that the latent period is 7 symbols long, and there is equal probability to encounter amino acids shown in Table 1 for each position of the period, e.g. letters A,R,N,D,C,Q,E,Z,X,W,Y,B are equiprobable at positions 1+7*N, other letters being absent at these positions. This table shows the artificial matrix used to represent the concept of latent periodicity. One of possible sequences satisfying the conditions in the table is:

AYEHSHCCBQKIDGZGRILSHQICHSHGYBNSDDIWNEWRCLZIFQAHHWYIWIDIQZ
RIIHQCXGMRNEEXECRGGCVIISHGZNEWRIEZILFNLEZKHMASCAGQTQGHNGH
WRCHZWISGLGEAQPZDERKAISDE

Let us assess the probabilities α (of meeting such, or better, homology between repeats by chance) and β (of obtaining the observed, or higher, value of mutual information by chance) for this sequence using the Monte-Carlo techniques. To do it we calculated the alignment score W from equations (1)-(3) using the affinity matrix BLOSUM50, as well as mutual information of the sequence (4) [26]. Then we generated 200 random shuffles of the sequence with the same amino acid composition. For each of these sequences, the values of similarity score W from (1) and mutual information were obtained. Then we calculated means and variances of both W and mutual information, and we obtained Z-scores using the formula (15). For our sample sequence, the calculated Z-values were equal to 2.96 and 6.5, respectively. That is, the probability α is roughly estimated as 0.05 and the probability β is roughly estimated as $10^{-9}$. The shown values of α and β illustrate that homology-based algorithms of searching for repeats are unable to identify such periodicities at statistically significant level.

**2.1 Analysis of Swiss-prot data bank.**

The concept of searching for latent periodicity with indels is graphically represented in Figure 1. It should be noted that the initial periodic profiles for iterative searching are taken from our database of latent periodicities found in Swiss-Prot using ID [24]. The information decomposition method had identified more than 12000 subsequences with significant periodicity of various lengths and types (we refer to amino acids in protein sequences as symbols and to amino acid sequences as symbolic sequences because information decomposition technique is regardless of what type of symbolic information is to process – nucleotides, amino acid residues or even written text). Nearly 20% of found amino acid sequences contained homologous periodicities, i.e. those periodicities had values of α less than $10^{-6}$, and homology search techniques are thereby enough sensitive to identify them.

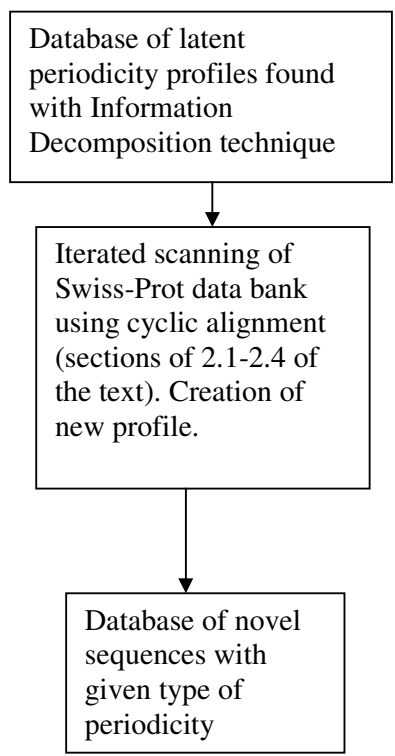

Fig.1 The framework of searching for latent periodicity with iterated profile analysis.

We excluded those cases from our analysis because RADAR [10] and REPRO [56] have been good tools to investigate them. We consider a few cases of homologous periodicity in this paper only to perform comparative analysis, while concentrating on non-homologous periodicities with $\alpha > 5*10^{-3}$ and $\beta < 10^{-6}$. RADAR and REPRO were also utilized in our analysis; they were unable to identify any periodicity in sequences of our interest.

Then these identified periodicities provided the initial data for searching for similar periodicities with allowed insertions and deletions. We chose the cases with latent periodicity that have the same period length and reside within functionally equivalent domains of different proteins. For these cases, we defined the latent periodicity matrices [24] that contain the numbers of amino acids at each position in the period. Finally, we merged all these matrices into one to derive the position-weight matrix (profile) from.

## 2.2. Noise decomposition

A profile **R** is a two-dimensional matrix, defining a weight of each symbol in each position. Profile comparison uses the mapping *[i, $b_j$] -> R* [27], where *i* is a position and $b_i$ is a symbol at position *i*. Profiles are usually obtained as a result of multiple alignment of protein sequences having some common property, and subsequently they are used to search for homologies in data banks [27]. However, we may consider the inverse problem: how to form a new profile from the results of profile analysis of data bank? It should have the following properties:

- sensitivity, i.e. it should be capable of finding at least a major part (ideally all) of related proteins;
- optimality, i.e. the score of these proteins should be as high as possible and statistical probability of casual finding of unrelated subsequence scoring this high should be as low as possible;
- specificity, i.e. we should find the least possible number (ideally no) proteins not having this property. The difference between b) and c) is that in the former case we deal with some statistical model of sequence data bank (usually in the form of a long sequence with given probabilities of symbols or groups of symbols), while in the latter case we we deal with real data bank. As we will see below, the difference between "optimality" and "specificity" is important.

If we create new profile from the results of searching a data bank and repeat this process, we can thereby perform iterative refinement of the initial profile. To make this possible, the iteration method should be asymptotically stable, i.e. after a certain number of iterations the new profile should be a good match to the previous one. This condition virtually defines the choice of formula to fill the position-weight matrix, since there is the theorem that specifies the connection between asymptotic frequencies of symbols in high-scoring segments (namely, the results of searching a data bank) and the scoring scheme [29]. It may be written in the form:

$$W_{i,j} = C \ln \frac{p_{i,j}}{f_i}, \tag{5}$$

where $W_{i,j}$ is an element of the position-weight matrix for symbol of type $i$ at position $j$, $p_{i,j}$ is fraction of symbol of type $i$ occurring at position $j$ within the high-scoring segments, and $f_i$ is the frequency of occurrence of symbols of type $i$ in scanned sequence, namely, in the data bank. Of course, pseudo-counts should be used in this formula where appropriate in order to avoid logarithm errors. The scaling parameter $C$ in the formula may be arbitrarily chosen (multiplying all weights and scores by a factor does not change the path of the alignment or its statistical significance, provided that indel costs are multiplied by the same factor), so we can choose it large enough to round the weight values to integers without substantial loss of precision in order to speed up the computation.

From the viewpoint of information theory, the problem of calculation of new profile from the results of profile analysis is similar to that of useful signal extraction from a noisy channel [30,31]. It is clear that certain patterns corresponding to frequent structural and functional features of proteins are over-represented (compared to what we could expect from their length and amino acid composition). Since profile searching is very sensitive, in most realistic cases some of the representatives of these structural units could be found even with a distantly related profile. If we include them in the multiple alignment to derive new profile, they will skew the profile into the direction of that over-represented group. It is likely to lead to a subsequent sharp increase of such representatives, and the initial motif would be completely changed [32]. If we are searching for common profile for a family of proteins or domains, it may result in unwanted effects known as "profile wandering". Moreover, it makes automation of successive iteration process virtually impossible – one has to choose "correct" scan results by hand in order to avoid profile "pollution". When the number of results runs into the hundreds and thousands, "time-consuming" turns into "impossible".

Information theory may be a way to circumvent this problem. The main idea of information-theoretic approach is to identify certain noise patters in noisy signal, and then to separate signal from noise being aware of these patterns. For example, if we observe the excess of noise at some frequencies, these frequencies should be most suppressed. This principle is really working in noise reduction systems. To set apart from "genetic noise", we propose the probabilistic consideration of sequence set that we call "noise decomposition" (ND). Let us represent this set as a mixture of "uncorrelated noise", composed of sequences having no homology to the proteins under investigation and characterized by "background" symbol probabilities $f_i$, and "correlated noise", composed of sequences generally unwanted but having sufficient level of homology to be found with profile analysis together with the desired sequences. For our purposes, the main difference of correlated noise is that the distribution of the numbers of symbol occurrences over different positions of profile alignment is not casual because we propose the existence of a pattern (or patterns) of these false positives. The resulting noise will apparently be position-specific, let us denote its distribution as $\pi_{i,j}$.

It is clear that correlated noise may be not of single type, but it may be some composition of a few different types. We assign a distribution to each of these types and call it $q^a_{i,j}$, where the upper index corresponds to different constituents of the correlated noise, composed of different types of false positives. Then the correlated noise can be expressed by this formula:

$$\pi_{i,j} = c_0 f_i + \sum_{k=1}^{N} c_k q^k_{i,j}, \quad \sum_{k=0}^{N} c_k = 1, \tag{6}$$

according to it, we rewrite the expression for calculation of the elements of the position-weight matrix (5) as:

$$\overline{W}_{i,j} = C \ln \frac{r_{i,j}}{\pi_{i,j}} \tag{7}$$

Here $c_k$ are "relative magnitudes" of different types of noise, in other words, their contribution into the resulting noise. The normalization in (6) is required to hold this condition:

$$\forall j, \sum_{\text{all } i} \pi_{i,j} = 1., \tag{8}$$

since $\pi_{i,j}$ are probabilities.

In order to define $c_i$, we will weight different types of noise as follows. Weighting coefficients $c_k$ should be proportional to the relative fraction of the corresponding type of noise in the total input signal, i.e. if the source bank contains $M_1$ sequences with correlated noise type 1, $M_2$ sequences with correlated noise type 2 and so on, and also $M_0$ sequences without correlation with our pattern, then the relative contribution of these noises into the resulting noise will be proportional to these numbers, i.e.

$$\frac{c_0}{M_0} = \frac{c_1}{M_1} = \frac{c_2}{M_2} = ... \tag{9}$$

However, in fact, the numbers $M_0$, $M_1$, $M_2$... are unknown. We will thereby estimate them from the results of initial search. Since the insignificant subsequences should be treated as noise, we will take into account only statistically significant results (see equation 15). If the search results contain $N_1$ sequences with correlated noise type 1, $N_2$ sequences with correlated noise type 2 and so on, we assume that

$$\frac{c_0}{N_0} = \frac{c_1}{N_1} = \frac{c_2}{N_2} = ... \tag{10}$$

Notice that $N_0$ could not be estimated this way (if we do that, $c_0$ will be set too low, and the new profile (7) will be distorted; our experiments demonstrated that the iterative process diverges under this condition). Our experiments show that 0.75 is a good estimate for $c_0$ in most cases (see later).

We calculated $r_{i,j}$ via pair-wise global alignment of all sequences in true positive set; let the alignment score of sequences $k$ and $l$ be $S(k,l)$. These values were utilized to calculate $T(k)$, which stands for prevalence of $k$-like sequences in true positive set:

$$T(k) = \sum_l \max(0, S(k,l) / \{\max(S(k,k), S(l,l))\}) \tag{11}$$

Index l runs through all the set of true positives. The meanings of terms in the sum (11) are: the term with $l = k$ always equals 1 (any sequence is self-similar), terms from unrelated sequences are zeroes, and terms from similar sequences range from 0 to 1. Therefore, we get $T(k) = 1$ in case that there are no sequences similar to k; we get $T(k) = N$ if all N sequences in the true positive set are identical; we get a value from 1 to N if sequences are similar, depending on similarity level. Then we calculate the values of $r_{i,j}$:

$$r_{i,j} = \sum_k p^k_{i,j} / T(k) \tag{12}$$

where $p_{i,j}$ are frequencies of occurrence of symbol i at position j in cyclic alignment of sequence k. Thus we eliminate possible over-representation of some sorts of sequences in Swiss-Prot data bank.

## 2.3 Iterated profile analysis.

Iterations were performed as follows. First of all, we used the initial matrix of periodicity determined via the algorithm described in Section 2.1, and we calculated the $W_{i,j}$ using (5). Then we used this position-weight matrix in the procedure of cyclic alignment, which will be described in the section 2.4 of the paper. Cyclic alignment of all amino acid sequences in Swiss-Prot was carried out, and all statistically significant results (Z>6.0) were gathered.

After that, we divided these results into classes, namely, true positives and false positives of different classes, for the formulae (6) and (7) to be applied. Two ways of division into classes were tried out, namely, keyword analysis and clustering.

We worked with the Swiss-Prot data bank, which is curated and contains some information about protein entries. We extracted information about identified proteins from their descriptions (DE field), keywords (KW field) and feature tables (FT field). We divided the results into three classes using keywords. Class 1 comprised proteins we wanted to identify, and classes 2, 3,4,… comprised groups of proteins we wanted to filter off. Then we considered the classes 1 as true positives and other classes as different types of correlated noise. This approach is applicable when proper information is presented in Swiss-Prot or any other database.

If such information is absent in a data bank, clustering may be used to split the proteins into classes. We performed the clustering experiment for protein kinases as follows. First, we made pairwise comparisons of identified subsequences using a global alignment method [39]. Then we built the distance matrix between those sequences using the formula:

Distance(A,B)=(AlignmentScore(A,A)+AlignmentScore(B,B))/2–AlignmentScore(A,B).

This distance matrix was used in single linkage cluster analysis, and threshold of merging was adjusted to output 2 large classes. Then we checked whether these clusters are related to 2 types of protein kinases, and we found incomplete correlation (about 90%), i.e. there were both serine-threonine and tyrosine kinases in each cluster, but the separation was about 90%. We concluded that the information in Swiss-Prot is more reliable than clustering, and it should be used when possible.

After formation of the classes, we calculated the values of $p_{i,j}$ and $q^k_{i,j}$, which are simply positional residue frequencies for the corresponding classes, as well as the values of $N_i$, the cardinalities of the classes. We let $f_i$ equal to the residue frequencies from Swiss-Prot. Then we were able to apply formulae (11), (10), (6) and (7) consecutively in order to obtain the new values of $W_{i,j}$, i.e. the new cyclic profile matrix.

Using the new $W_{i,j}$ we repeated the search of Swiss-prot data bank according to Section 2.4 and we obtained the new set of amino acid sequences with Z>6.0. After that, we repeated theprocedure of selection of true and false positives as shown above in this section, and we recalculated $W_{i,j}$ again. The iteration process was carried out until the set of results after iteration was virtually the same as the set before the iteration. It means that similarity of two sets was more 95%. Our experiments showed that 3 to 5 iterations were enough in most cases.

## 2.4 Cyclic alignment and statistical significance.

It is clear that noise decomposition is quite applicable to conventional (linear) profiles [27], but in the present study it was used for training of periodic profiles in order to investigate latent periodicity of protein sequences [24, 26]. The reasoning given above is valid and expressions (5)-(12) hold in both cases. The problem of finding a good cyclic profile is much more sophisticated than the problem of finding of linear profile, since there are a number of diverged copies (repeats) of a pattern within a sequence, instead of one. Internal divergence of repeats superimposes on divergence between sequences; hence, cyclic patterns are much more feebly marked and hard to correlate to some structural or functional unit. We have made a modification of the algorithm in [33], which is called locally-optimal cyclic alignment [57].

What is cyclic alignment? The conventional alignment is matching of a sequence to some pattern, for example, "QWERTY". The cyclic alignment is then matching of that sequence to virtual periodically-elongated pattern "…QWERTYQWERTYQWERTYQWERTY…"; of course, in most cases only a part of the sequence will be matched (local cyclic alignment). Our main idea is to present cyclic alignment in the form of a path that connects the nodes of a two-dimensional cylindrical lattice, where one of the coordinates corresponds to position in (linear) sequence, and another (cyclic) is for position in the cyclic profile (compare to conventional sequence alignment, which can be presented in the form of a path between nodes of the two-dimensional lattice, coordinates being the positions in compared sequences). This path contains diagonal steps, which describe matching of a symbol from the sequence and a position of the profile, as well as steps along the axes, which describe insertions or deletions. Every such path has a total score, which is the sum of gap penalties and weights of symbol-to-position matches (see [57] for details and figures).

The optimal cyclic alignment is the path with the highest possible total score. We have shown that it can be found when we fill cell-by-cell the similarity matrix $S_{i,j}$ with one of its indices (*i*) being cyclic or wrapped, namely, $S_{i-L,j} \equiv S_{i+L,j} \equiv S_{i+2L,j} \equiv … \equiv S_{i,j}$, where *L* is the period length [57], similarly to finding of the best linear alignment using Smith-Waterman formula [28]. The formulae for recursive filling of $S_{i,j}$ we found are:

$$S_{i,j} = \max\{S'_{i,j}, \max_{1 \leq k \leq L-1}[S'_{(i-k),j} - d_k]\}, \text{ where} \qquad (13)$$

$$S'_{i,j} = \max\left\{0, S_{i-1,j-1} + w_{i,j}, \max_{1 \leq k \leq j}\left[S_{i,j-k} - d_k\right]\right\} \qquad (14)$$

Here $w_{i,j}$ is the weight of the *j*th symbol in the sequence at the *i*th position in the profile (it is element of matrix *W* from (5) or (7)), $d_k$ is the gap penalty for insertion/deletion of *k* successive symbols. As usual, to find the optimal local alignment, we have to find the highest element of S-matrix and recreate the path down to first element equal to zero. The value of the highest element is the total score of optimal local alignment; this value was used to check whether the alignment is statistically significant.

The Monte-Carlo technique was used to assess the statistical significance of alignments [58]. The assessment was performed separately for each sequence, taking into account its length and composition. To assess the statistical significance of an alignment in this study, we aligned a number of random sequences with the same length and composition as the real sequence (to avoid composition bias effects) against the same cyclic profile. Then we could estimate the Z-value of the obtained alignment from the mean and the variance of those random scores, assuming that the distribution of obtained weights is normal [34, 35]:

$$Z = \frac{S_{real} - E(S)}{\sqrt{D(S)}} \qquad (15)$$

The threshold value of Z was chosen to be equal to 6.0. Our numerical experiments showed that we are unlikely to observe Z-values greater than 6.0 when analyzing a random test sequence with the same number of symbols as the total number of symbols in Swiss-Prot.

The described algorithms were implemented in C++, and are available upon request by e-mail (for details visit our Web site http://bioinf.narod.ru/periodicity).

### 3. Results

First of all, we wanted to show the usefulness of our techniques to search for known types of tandem repeats. Ankyrin and leucine-rich repeats were chosen for this purpose. The initial profiles of these repeats were obtained using ID with the period lengths equal to 33 and 24 residues, correspondingly. The results are shown in Table 5.

We identified 146 of 150 sequences containing at least 3 marked ankyrin repeats (3 times the period length was chosen to be minimal length of latently-periodic subsequences). Needless to say, identified periodic subsequences covered the marked repeats. Scanning the rest of Swiss-Prot, we identified additional 57 sequences, in which less than 3 ankyrin repeats were marked by conventional techniques. Hence, we identify more repeats than conventional techniques are able to identify. The only false positive, hypothetical protein P50938 was identified because of perfect tandem periodicity of another type in it.

We also identified 261 of 270 sequences with leucine-rich repeats with no false positives. As in the previous case, we identified many additional leucine-rich repeats in those sequences. For example, in protein P09661 (U2 small nuclear ribonucleoprotein) we identified additional 4th repeat, which is unseen by conventional techniques (we tested that with PFam, SMART, as well as with dedicated repeat-finding software, namely, REP [11], REPRO [56] and RADAR [10]). The 3D-structure of this protein is known, and the presence of the 4th repeat is evident. Another example is the protein P16473 (thyroid stimulating hormone receptor), where we identified 10 leucine-rich repeats, while other techniques identify no more than 6 repeats. The modeled 3D-structure of this protein testifies that the identified region consists completely of LRR repeats. Hence, the usage of ID and cyclic alignment techniques may lead to new discoveries even in well-known repeat families. We also identified additional reliable repeats in many other known repeat families, but we are not concentrated on those studies (the reason is explained in Section 2.1).

Table 2. Protein kinases where latent periodicity without insertions and deletions was firstly found

| Swiss-Prot ID | Start position | End position | Protein description in Swiss-Prot |
| --- | --- | --- | --- |
| KDBE_SCHPO | 400 | 478 | Putative serine/threonine-protein kinase C22E12.14C (EC 2.7.1.). |
| KEMK_MOUSE | 85 | 181 | Putative serine/threonine-protein kinase EMK (EC 2.7.1.). |
| KPC1_ASPNG | 954 | 1056 | Protein kinase C-LIKE (EC 2.7.1.). |
| KPCL_RAT | 526 | 565 | Protein kinase C, ETA type 2.7.1.-) (NPKC-ETA) (PKC-L). |
| M3KA_HUMAN | 97 | 181 | Mitogen-activated protein kinase 10 (EC 2.7.1.) (mixed lineage kinase 2) (protein kinase MST). |
| CC22_XENLA | 85 | 148 | Cell division control protein 2 homolog 2 (EC 2.7.1.-) (P34 proteinkinase). |
| CC2_CARAU | 88 | 160 | Cell division control protein 2 homolog (EC 2.7.1.-) (P34 proteinkinase) (cyclin-dependent kinase 1) (CDK1). |

In our present study, we identified 7 protein kinase amino acid sequences with latent periodicity with the period length equal to 18 residues, in the absence of insertions and deletions (Table 2, Fig.2). All of them were serine-threonine protein kinases except

M3KA_HUMAN having dual specificity. The periodicities were located in the protein kinase catalytic domains (in a number of cases we also encountered periodicity with period lengths of multiples or divisors of 18). This suggested that the periodicity of 18 amino acids is a characteristic property of protein kinase active sites. We had set a problem to find out, first, to what extent this periodicity is peculiar to protein kinase active sites if insertions and deletions are permitted, and, second, whether serine-threonine and tyrosine protein kinases share the same periodicity pattern, and whether we can improve sensitivity and specificity of search by using of separate periodicity profiles for different kinase types instead of one.

We averaged amino acid occurrence frequencies in different profile positions of all 7 cases mentioned above and made the initial position-weight matrix using formula (2). This matrix was used to scan Swiss-Prot release 41 [36] using cyclic profile alignment. A number of values of gap opening and extension costs were tried out, taking into account sensitivity and specificity. Values of 3.8 for gap opening and 0.7 for gap extension were found to be optimal and used thereafter.

As a result of this initial scanning, we obtained about 100 statistically significant periodic subsequences from both serine-threonine and tyrosine protein kinases. It is a fact that catalytic domains of serine-threonine and tyrosine protein kinases have fairly homologous primary sequences as well as similar 3D structure [37]. Hence, we decided to form two periodic profiles according to these two types of protein kinase catalytic cores. To do it we divided results into classes and formed two new position-weight matrices using (6)-(12), so that in one case serine-threonine protein kinases were considered to be true positives and tyrosine protein kinases were considered to be a kind of correlated noise, and vice versa for the other matrix. Noise decomposition resulted in 2 position-weight matrices, which were later optimized (trained) with iterative searches in order to find the highest number of serine-threonine or tyrosine protein kinases, respectively, while keeping the specificity. We performed iterations until the profile was nearly stable (generally 3-5 iterations were enough).

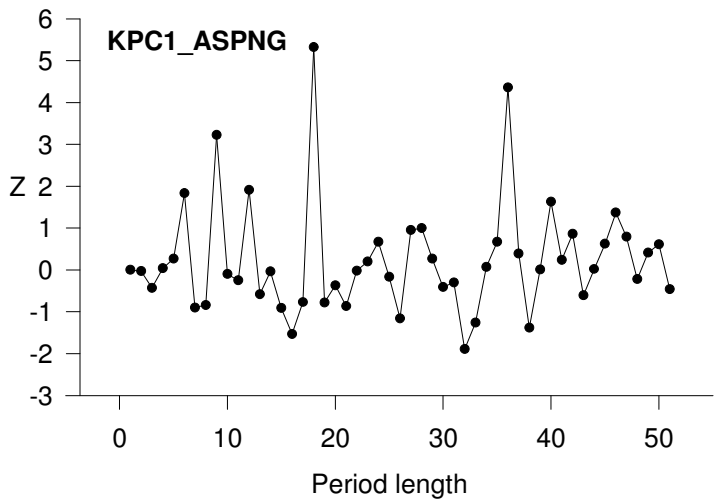

Fig. 2. Information decomposition spectrum for amino acid sequence from KPC1_ASPNG (residues 954 – 1056). Full protein kinase domain of this protein contains residues 770-1030.

```
789abcdefghi123456789abcdefghi123456789abcdefghi123456789abcdefghi123456789
abcdefghi123456789abcdefghi1
WWAFGVLIYQMLLQQSPFRGEDEDEIYDAILADEPLYPIHMPRDSVSILQKLLTREPELRLGSGPTDAQEVMSHA
FFRNINWDDIYHKRVPPPFLPQISSPTD
```

Upper sequence shows the positions of the 18 amino acid latent periodicity. The matrix of 18 amino acid periodicity for serine-threonine protein kinases is shown on the site http://bioinf.narod.ru/periodicity. The number 1 shows the first column of the matrix, number 2 shows the second position of the matrix and so on. Lower sequences is the sequences from KPC1_ASPNG.

Table 3. Results of the analysis of Swiss-prot (release 41) data bank for presence of amino sequences with the similarity to the matrix with latent periodicity.

| Matrix type | Serine/threonine-protein kinases | Tyrosine-protein kinases |
| --- | --- | --- |
| Total number of protein kinases present in Swiss-Prot release 41 | **1116** (62 of them are dual-specificity kinases) | **348** |
| Total number of protein kinases having Z more than 6.0 | **903** | **312** |
| False positives | **37** | **11** |
| Another type protein kinases found among false positives | **4** (tyrosine kinases) + **1** (unknown kinase) | **4** (serine/threonine kinases) + **6** (tyr kinase-like |

Table 6 represents the process of noise decomposition. One may see from the table that, although training of the initial profile using test set of serine/threonine protein kinase sequences raised the number of identified S/T kinases by nearly 10 times, the accuracy of distinction between the two types of kinases is as low as 5.64. Introduction of the mixing coefficient (6) greatly decreases the number of identified tyrosine kinases, but higher values of this coefficient vastly distort the profile and affect its sensitivity and specificity. As it was said above in the text, the best value of $c_c$ is 0.75 or thereby; it increases separation between the two types of kinases to 264.

The summary of final results is presented in Table 3. In both cases, we found latent periodicity in more than 80% of the target protein kinases, reached 96% specificity and 99% kinase separation. The examples of alignments are shown in Table 4. To obtain the latest versions of latent periodicity profiles and sets of results for these and other investigated instances, visit the authors' Web site at http://bioinf.narod.ru/periodicity.

Table 4 Examples of periodical alignments for two serine-threonine and two tyrosine protein kinases. Alignment upper sequence shows the positions of the 18 amino acid period and lower sequence is amino acid sequence with latent periodicity. Corresponding matrixes of latent periodicity are shown on the site http://bioinf.narod.ru/periodicity.

| Protein description | Accession number | Z | Position in protein | Alignment |
|---|---|---|---|---|
| Serine/threonine-protein kinase CBK1 | P53894 | 10.7 | 449-610 | `23456789abcdefg--------hi123456789abcdefghi123456-`<br>`DVTRFYMA-ECILAIETIHKLGFIHRDIKPDNILIDIRGHIKLSDFGLST`<br><br>`---789abcdefghi------12--345----6789abcdefghi12345`<br>`GFHKTH-DSNYYKKLLQQDEATNGISKPGTYNANTTDTANKRQTMV---V`<br><br>`6789abcdefghi123456789abcdefghi123--456789abcdefgh`<br>`DSISLTMSNRQQI--QT--WR---KSRRLMAYSTVGTPDYIAPEIFLYQ-`<br><br>`i123456789abcdefghi123456`<br>`GYGQECDWWSLGAIMYECLIGWPPF` |
| Serine/threonine-protein kinase, cGMP-dependent protein kinase, isozyme 1 | Q03042 | 16.4 | 505-655 | `cdefghi123456789abcdefghi123456789ab-cdefg---hi123`<br>`SERHIMLSSRSPFI----CRLYRTFRDEKYVYMLLEACMGGEIWTMLRDR`<br><br>`45--6789abcdefghi123456789ab--cdefghi123456789abcd`<br>`GSFEDNAA----QFIIGCVL----QAFEYLHARGIIYRDLKPENLMLDER`<br><br>`efghi123456789abcdefghi123456789abcdefghi123456789`<br>`GYVKIVDFGFAKQIGTSSKTWTFCG-TPEYVAPEIILNKGH-DRAVDYWA`<br><br>`abcdefghi123456`<br>`LGILIHELLNGTPPF` |
| NT-3 growth factor receptor precursor, TrkC | Q16288 | 16.3 | 663-764 | `56789abcdefghi123456789abcdefghi123456789abcdefghi`<br>`ASGMVYLASQHFVHRDLATRNCLVGANLLVKIGDFGMSRDVYSTDYYRLF`<br><br>`--1--23456789abcdefg-hi12--3456789abcdefghi1234567` |

| tyrosine DE kinase | | | | `NPSGNDF----CI-----WCEVGGHTMLPIRWMPPESIMYRKFTTE-SDV`<br><br>`89abcdefghi1`<br>`WSFGVILWEIFT` |
| --- | --- | --- | --- | --- |
| Tyrosine-protein kinase transforming protein YES | P00527 | 16.4 | 370-457 | `56789abcd-efghi123456789abcdefghi123456789abcd-efg`<br>`ADGMAY-IERMNYIHRDLRAANILVGDNLVCKIADFGLAR-LIEDNEYTA`<br><br>`h-i123456789abcdefghi1-23456789abcdefghi12`<br>`RQGAKFPIKWTAPEAALYGRFTIK--SDVWSFGILLTELVTK` |

After investigation of latent periodicity of protein kinases, we started this analysis for other protein families. We should say that not 100% of latent periodicities identified in protein sequences with ID technique turn out to be peculiar to whole families. In fact, about 70% of them reveal similar proteins during subsequent iterative searches. In the remaining 30% of cases, the latent periodicity was peculiar only for single sequence, where it was initially found, or it was identified in a group of amino acid sequences of different function. Such results show that these latent periodicities are not peculiar to the families they were found in.

Further application of cyclic profile analysis to latent periodicities, identified in protein sequences with ID technique, allowed us to identify other protein families having latent periodicity. In this study, we demonstrate the presence of latent periodicity in 22 other families. Latent periodicity is identified in more than 75% of members of each family. The list of these families is shown in Table 5. The corresponding profiles are presented at http://bioinf.narod.ru/periodicity/new

Table.5 Protein families where latent periodicity with insertions and deletions were found. The first 2 cases are examples of homologous periodicity identified with our techniques. Cyclic alignments with corresponding cyclic profiles are shown at the web site http://bioinf.narod.ru/periodicity/new

| N | Protein family | Period length, aa : | Number of found proteins with latent periodicity |
| --- | --- | --- | --- |
| 1 | Ankyrin repeats | 33 | 203 |
| 2 | Leucine-rich repeats | 24 | 261 |
| 3 | Acetolactate synthases | 25 | 28 |
| 4 | Granule-bound glycogen synthase | 41 | 11 |
| 5 | Trna synthetases | 17 | 71 |
| 6 | Acyl and adenyl transferases, EIF | 6 | 136 |
| 7 | Potassium channels | 25 | 55 |
| 8 | Acyl-transferating synthases, active site | 25 | 124 |
| 9 | Dethiobiotin synthases | 30 | 15 |
| 10 | Various GTP-binding proteins | 23 | 270 |
| 11 | Tryptophan 2-monooxygenases | 28 | 6 |
| 12 | Various CoA-related proteins, maybe CoA- | 30 | 42 |

| | binding site | | |
|---|---|---|---|
| 13 | Asparatyl proteases | 41 | 54 |
| 14 | MHC I Antigens | 12 | 148 |
| 15 | Methyltransferases | 31 | 30 |
| 16 | ACC oxidases | 32 | 17 |
| 17 | Polymerase core | 13 | 103 |
| 18 | Interleukin-12, growth factors | 14 | 12 |
| 19 | ATP synthases, active site | 14 | 82 |
| 20 | Hemagglutins | 15 | 92 |
| 21 | Hedgehog proteins | 16 | 59 |
| 22 | Actins | 17 | 256 |
| 23 | Lyases | 10 | 428 |
| 24 | Pyridoxal phosphate binding proteins | 16 | 76 |

Table 6. Noise decomposition of protein kinase superfamily. Results of searching with various profike matrices are shown; the profille was to be optimized for serine/threonine kinases). Different choices of mixing coefficient $c_1$ (here $c_0=1-c_1$) lead to different results of searching. The value of 0.25 is optimal for recognition of protein kinases.

| Profile | S/T kinases | Y kinases | Other proteins |
|---|---|---|---|
| Initial profile found with ID | 63 | 41 | 18 |
| Initial profile, after 5 iterations with test set of S/T kinases, no noise decomposition performed | 615 | 109 | 58 |
| Initial profile, after noise decomposition and 1 iteration: | | | |
| c1 = 0.1 | 554 | 29 | 69 |
| c1 = 0.25 | 528 | 2 | 61 |
| c1 = 0.5 | 407 | 0 | 80 |
| c1 = 0.75 | 229 | 0 | 99 |
| c1 = 1 | 101 | 0 | 214 |
| c1 = 0.25, after 5 iterations | 866 | 4 | 33 |

We also counted the numbers of latently-periodic Swiss-Prot sequences, which did not belong to corresponding families. These numbers did not exceed 7% of the numbers of proteins in corresponding families, and they were equal to zero for two thirds of them. We propose that the presence of such sequences may be caused by imperfect annotation of Swiss-Prot data bank and possible additional functions of hypothetical proteins.

## 4. Discussion

The latent periodicity notion and search technique was initially presented in [22] and refined in subsequent works [23-26]. As a result of performed studies, we revealed the existence of various types of latent periodicity in numerous amino acid sequences [24,26]. These found latently periodic sequences belonged to various proteins, with various period lengths and periodicity patterns. However, the question of functional significance of identified latent periodicities and its correlation with protein structures remained open. To a great extent this resulted from the information decomposition method being incapable of revealing latent periodicity interrupted with insertions and deletions, so this approach omitted a substantial subset of proteins with certain functional domains, so that no inference about the relationship between latent periodicity and protein functionality could be made. This paper presents a pioneer work that shows the existence of latent periodicity for a whole protein families. We have achieved this result by means of the development of iterated mathematical method and software, capable of finding statistically significant latent periodicity of a predefined type (found by ID method) in presence of insertions and deletions.

The applied noise decomposition technique and the iterative analysis do not destroy the latent periodicity we identified in protein kinases and other protein families. This can be seen from Figure 3, in which the information decompositions of a few protein kinase amino acid sequences after noise decomposition and iterative analysis are presented. The periodicity of 18 residues is evident from the spectrograms of information decompositions. Similar decomposition spectra may be obtained from all the other protein kinase amino acid sequences (as well as in sequences from other protein families aligned with corresponding profiles) after noise decomposition and iterative analysis.

Let us discuss the existing limitations of our current abilities to identify periodicities. First, it should be noted that the noise decomposition technique was in the present study used in conjunction with the information decomposition technique, which provided the initial periodicity profiles (see Section 2.1). Hence, it is capable of identifying the (both latent and explicit) periodicity with indels, but the types of periodicity should be identifiable without allowance of indels. So we may omit some homologous periodicities, in which no sequences of their types in Swiss-Prot are free of indels. Nevertheless, searching for homologous tandem repeats is rather investigated problem, and it was not the aim of our investigations. We concentrated on searching for feebly marked periodicities, and the conjunction of ID and ND techniques is the only method to study them so far. One may also notice that the noise decomposition method itself can be applied to study any profile. From this viewpoint, it may be considered as independent technique.

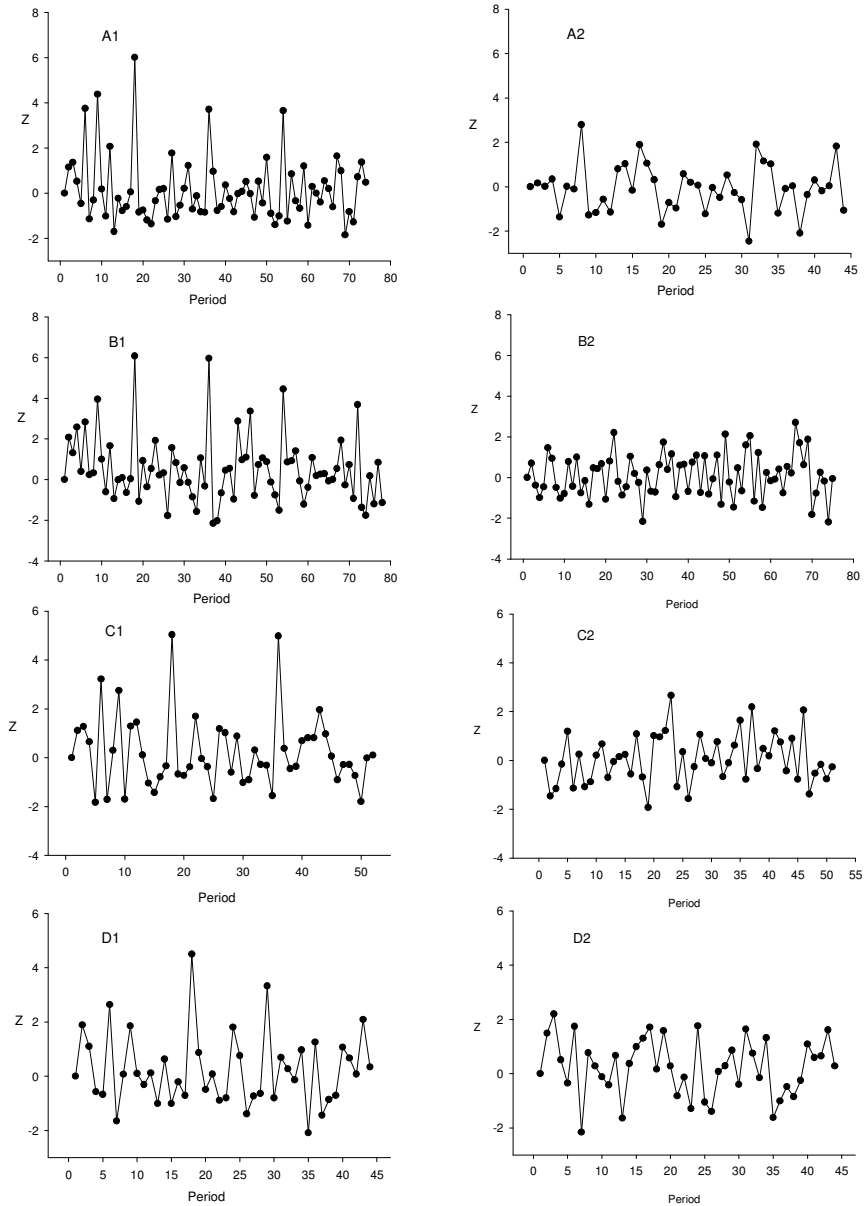

Fig.3. Information decomposition of serine-threonine and tyrosine protein kinases after noise decomposition and iterration procedures (number 1 in the lebel) and before it (number 2 in lebel). The corresponding sequences after cyclic alignment are shown in Table 4.
A- P53894; B - Q03042; C - Q16288; D - P00527

Second, the necessity to find initial periodicity types without insertions and deletions also affected our choice of maximal possible size of indels; it was limited to the period length. If we allow for larger deletions, we would find many linked dispersed repeats in different parts of sequences instead of periodicity in domains, which is of our interest. It is possible that there are proteins in some investigated families, in which such large indels occurred nevertheless; in that case, we were unable to identify them. But we did that for the sake of specificity.

Third, our numerical experiments have demonstrated that the used estimation of statistical significance (15) is valid only if length of alignment is greater than 50 [58]. It means that the cyclic profile search technique is capable of identifying periodicity, which ranges along 50 or more amino acid residues; short dispersed repeats would not be identified. But it was mentioned above that we avoided dispersed repeats in our searches. Moreover, the ID technique can be good at finding periodicities only if at least 3 consecutive repeats are present, and we developed the ND technique with this limitation of minimal length of latently-periodic subsequences (i.e. 54 residues for protein kinases). That is, if only a part of protein kinase catalytic domain is conserved, our software would not identify it; however, such scrap of the domain is unlikely to be functioning.

We have made the comparative analysis of the power of our method, in which we tried to reveal the periodicity of protein kinases and other latently-periodic protein families (Table 5) with RADAR [10] and REPRO [56] software. Both failed to do that.

In this paper, we did not pose a problem to reveal all existing cases of latent periodicity in protein families. It is rather laborious problem, and it requires great computational power. We focused our efforts to demonstrate that we are able to identify periodicity in most of proteins in families, in which only a few members were found to be latently-periodic. The periodic structure of all these proteins and domains was not described before.

Let us discuss specifically the latent periodicity of protein kinases, its possible meaning and relationship with the spatial organization of proteins. Protein kinases, i.e. enzymes whose function is to transfer phosphate residues from ATP to other proteins, are known to provide a important role in cell signaling. There are many subfamilies of protein kinases with internal homology of 90% and higher. Mutual homology between subfamilies is much weaker, usually about 30%. Two classes of protein kinases are structurally very similar, serine-threonine kinases and tyrosine kinases (according to the phosphorylated residue). In addition, there also exist protein kinases with dual specificity [40]. There are also other types of protein kinases structurally dissimilar to these kinases, which phosphorylate other residues. They do not share the types of periodicity described in this paper, nor were they found to have any other common periodicity pattern. We think they could have another type of latent periodicity or the presence of deletions and insertions of big sizes in amino acid sequences could disturb to find of latent periodicity in protein kinases which are differ from serine-threonine and tyrosine kinases by noise decomposition method.

As is well-known [41,42], the catalytic domain of protein kinases, where our periodic subsequences reside, could be divided into 12 subdomains, on the one hand, highly evolutionarily conservative and, on the other hand, related to 3D-structure elements [43,44]. Subdomains I-IV are responsive for ATP binding and form antiparallel β-sheets. Subdomains VIa-XI bind the substrate and initiate the phosphate ion transfer. They are a bit more variable (presumably to provide substrate specificity) and composed of mostly α-helices.

Subdomains alternate with less conservative sites that usually form loops, with the period of these alternations being close to 18 residues. Using our protein kinase periodicity profiles, in various protein kinases we find periodic sites, about 100 residues long, located in subdomains VIb, VII, VIII and IX. These subdomains contain functionally important features such as the catalytically active asparatic acid residue in subdomain VIb and the activation loop between subdomains VII and VIII. Many amino acid residues within these subdomains are of critical importance for proper folding and functioning of the active center. These are the asparatic acid residue mentioned above, the valine residue that interacts with ATP adenine, the lysine residue that interacts with phosphate ion, the asparagine and asparatic acid residues in subdomain VII that retain inhibiting and activating Mg ions, the asparatic acid residue in subdomain IX that stabilizes the catalytic loop, and also a few other residues that provide ionic bonds and regulate enzyme activity, being subject for phosphorylation or autophosphorylation [45]. Notice that, for example, the aforementioned asparatic acid residues are separated from each other by 18 and 36 residues (the numbers are given for cAMP-dependent mammalian protein kinase A, which is usually a model enzyme for studying kinase structure; we obtained similar values for other proteins), so they are both placed at the same position in the periodic repeat (namely, position 2), although their functions are different. Therefore, we see that the period length is close to a subdomain. To make sure we compared cyclic profile alignment of protein kinase A (Swiss-Prot accession P05132) with its subdomain structure. It turned out that subdomain borders are located at period positions 14 and 15. Hence, there is a clear relationship between periods and subdomains.

It was previously proposed [46,47] that tyrosine protein kinases were evolutionarily derived from serine-threonine ones by means of isolation of catalytic domain nucleotide sequences by insertion of introns and subsequent pasting of these mobile elements, slightly altered with mutations, into some other proteins with kinase activity, called "ancestral kinases". This would result in greater variability of catalytic domain lengths in tyrosine kinases, because there are no gap restrictions for mobile elements. Our results favor that hypothesis, because we found out that insertions and deletions occur almost 2 times more frequently in tyrosine kinases than in serine-threonine ones (on average, 5.96 vs 3.05 insertions and deletions per site), i.e. we observe larger deviations from perfect periodicity.

At this time the exact origin of observed latent periodicity shown in this work is not clear, however, we can propose a few possible explanations of this phenomenon. First, we may suppose that catalytic domains were initially much smaller than what we observe now. However, they were able to duplicate and duplications were properly arranged to form even more catalytically active domain. It is a fact that DNA sequence repeats facilitate replication errors at their location, thereby promoting new tandem repeats. We suppose that as the number of repeats grew, the ancestor protein benefited, i.e. its catalytic activity and structure stability increased. It also means that periodical structures could extend over modern three dimensional structure limits. Subsequent mutations formed even better packed structure of these domains and fine-tuned the functionality, at the same time the mutations resulted in periodicity diffusion, loss of homology between distinct repeats. That means that we may call the residual internal similarities "echoes" of the ancient protein formation processes [48-50].

Latent periodicity may be also involved in stabilization of protein structure and proper folding. It is well known that protein folding is supervised by chaperone proteins that bind to growing polypeptide chain [51,52]. This binding is not strictly specific but there are certain binding preferences, the main factors being charge and hydrophobicity of amino acid sequence sites [53,54]. We suppose that periodic distribution of these properties along the sequence facilitates uniform distribution of chaperones and such uniformity is required (or desirable) for fast and proper folding. For many cases we observe structure-related periodicity, that is, different positions in a period correspond to different secondary structure preferences. For example, a period may consist of two parts, one showing α-helix preference and another showing β-structure preference [57]. At that, the periodic motif itself determines a supersecondary structure peculiar to a type of protein domains or single-domain proteins.

What is the extent of prevalence of latent periodicity, is it a common property of structural and functional protein units? This question could be answered only after building of complete database of latent periodicity profiles and their structural and functional features. Its creation is likely to demand great processing power and development of new methods for latent periodicity detection and characterization., Our up-to-date results show that 25 different protein families have latent periodicity, and number of these families is increasing each week as our investigations go on. We can propose that latent periodicity is common phenomenon for great number of protein families.

**References**

1. Benson DA, Karsch-Mizrachi I, Lipman DJ, Ostell J, Rapp BA, Wheeler DL (2000) Nucleic Acids Res 28: 15-18

2. Stoesser G, et.al. (2001) Nucl. Acids Res 29:17-21

3. Adams MD, et.al. (2000) Science 287: 2185-2195

4. Venter JC, et.al. (2001) Science 291: 1304-1351

5. Heringa J (1994) Comput Chem 18: 233-243

6. Heringa J Argos P (1993) Proteins 17: 391-410

7. Heringa J (1998) Curr Opin Struct Biol, 8: 338-345

8. Benson G (1997) J Comput Biol 4: 351-367

9. Benson G (1999) Nucleic Acids Res 27: 573-580

10. Heger A, Holm L (2000) Proteins 41: 224-237 (2000).

11. Andrade MA, Ponting CP, Gibson TJ, Bork P (2000) J Mol Biol 298: 521-537

12. Taylor WR, Heringa J, Baud F, Flores TP (2002) Protein Eng. 15: 79-89

13. Lobzin VV Chechetkin VR (2000) Uspekhi Fizicheskikh Nauk 170: 57-81

14. Dodin G, Vandergheynst P, Levoir P, Cordier C, Marcourt P (2000) J Theor Biol 206: 323-326

15. Jackson JH, George R, Herring PA (2000) Biochem. Biophys. Res. Commun, 268: 289-292

16. Rackovsky S (1998) Proc. Nat. Acad. Sci., 95: 8580-8584

17. Chechetkin VR, Lobzin VV (1998) J Biomol Struct Dyn, 15: 937-947

18. Coward E, Drablos F (1998) Bioinformatics, 14: 498-507

19. Voss RF (1992) Phys. Rev. Lett 25: 3805-3808

20. Silverman BD, Linsker R (1996) J.Theor.Biol., 118: 295-300

21. McLachlan AD, J.Phys.Chem, 97: 3000 (1993).

22. Korotkov EV Korotkova MA, (1995) DNA Seq, 5: 353-358

23. Korotkov EV Korotkova MA, Tulko, JS (1997) CABIOS, 13: 37-44

24. Korotkova MA, Korotkov EV Rudenko VM (1999) J. Mol. Model., 5: 103-115

25. Chaley MB, Korotkov EV, Skryabin KG (1999) DNA Res., 6: 153-163

26. Korotkov EV, Korotkova MA, Kudryshov NA (2003) Phys Let. A, 312: 198-210

27. Gribskov M, McLachlan AD, Eisenberg DB (1987) Proc. Natl Acad. Sci. USA, 84: 4355-4358.

28. Smith TF, Waterman MS, (1981) J. Mol. Biol., 147: 195-197

29. Karlin S, Dembo A, Kawabata T (1990) Ann. Stat. 18: 571-581

30. Schmidt JP (1998) Pac. Symp. Biocomput., 561-572

31. Wilbur WJ, Neuwald AF (2000) Computers&Chemistry, 24: 33-42

32. S.F.Altschul SF, Koonin EV (1998) TIBS, 23: 444-447

33. Fischetti V, Landau G, Schmidt J, Sellers P (1992) In Proceedings of the 3rd annual Symposium on Combinatorial Pattern Matching, Eds. Apostolico A., Crochemore M., Galil Z., Manber U., Lecture Notes in Computer Science volume 644, Springer-Verlag, p.111-120

34. Webber C, Barton GJ (2001) Bioinformatics, 17: 1158-1167

35. Chaley MB, Korotkov EV Kudryashov NA (2003) DNA Sequence, 14: 37-52

36. Bairoch A, Apweiler R (2000) Nucleic Acids Res., 25: 45-48

37. Taylor SS, Radzio-Andzelm E, Hunter T (1995) FASEB J. 9, 1255-1266.

38. V.L.Junker VL, Apweiler R, Bairoch A (1999) Bioinformatics, 15: 1066-1067

39. Needleman SB, Wunsch CD (1970) J. Mol. Biol., 48: 443-453

40. Kentrup H et.al., J Biol Chem., 271: 3488-3495

41. Hanks SK, Quinn AM, Hunter T (1988) Science, 241: 42-52

42. Hunter T (1991) Methods Enzymol., 200: 3-37

43. Taylor SS, Knighton DR, Zheng J, Ten Eyck LF, Sowadski JM (1992) Annu.Rev.Cell Biol. 8: 429-462

44. Goldsmith EJ, Cobb MH (1994) Curr.Opinion Struct.Biol. 4: 833-840

45. Taylor SS, Radzio-Andzelm E (1994) Structure, 2: 345-355

46. Kruse M, Muller IM, Muller WE (1997) Mol.Biol.Evol., 14, 1326-1334

47. Muller WE, Kruse M, Blumbach B, Skorokhod A, Muller MI (1999) Gene, 238: 179-193

48. Ohno S (1970) Evolution by gene duplication. Springer-Verlag, Berlin

49. Ohno S, Epplen J.T. (1983) Proc. Natl. Acad. Sci. USA, 80: 3391-3395

50. Ohno S (1984) J. Mol. Evol., 20: 313-321

51. Ruddon RW, Bedows E (1997) J.Biol.Chem., 272: 3125-3128

52. Thulasiraman V, Yang CF, Frydman J, (1999) EMBO J. 18: 85-95

53. Knarr G, Modrow S, Todd A, Gething MJ, Buchner J (1999) J.Biol.Chem., 274: 29850-29857

54. Takenaka IM et.al. (1995) J.Biol.Chem., 270: 19839-19844

55. Karlin S, Altschul F (1990) Proc. Natl. Acad. Sci. USA, 87: 2264-2268.

56. George RA, Heringa J (2000) Trends Biochem. Sci., 25:515-517.

57. Laskin AA, Chaley MB, Korotkov EV, Kudryshov NA (2003) Mol. Biol. (Russian), 37:663-674.

58. Chaley MB, Korotkov EV, Kudryashov NA. (2003) DNA Sequence, 14:37-52.